\documentclass[journal,twocolumn,singlespace]{IEEEtran}
\usepackage{amsbsy,amsmath,amssymb,epsfig,bbm,mathrsfs,multirow,amsthm,subcaption,caption,graphicx,epstopdf,multicol,lipsum,float}

\usepackage{adjustbox}
\usepackage{multicol}
\usepackage{enumerate}
\usepackage{tikz}
\usepackage[hidelinks]{hyperref}

\newenvironment{myalign}{\small\par\nobreak\noindent\align}{\endalign}
\usepackage{enumitem}

\usepackage{algorithm}
\usepackage{algorithmic}

\setlist[itemize]{leftmargin=*}

\begin{document}

\title{Multi-Objective Optimization for Drone Delivery}

\author{\IEEEauthorblockN
{Suttinee Sawadsitang\IEEEauthorrefmark{1},
 Dusit Niyato\IEEEauthorrefmark{1}},
 Puay Siew Tan\IEEEauthorrefmark{2},
 Ping Wang\IEEEauthorrefmark{3}, 
 Sarana Nutanong\IEEEauthorrefmark{4} \\
\IEEEauthorblockA{
\IEEEauthorrefmark{1} School of Computer Science and Engineering, Nanyang Technological University\\
\IEEEauthorrefmark{2}Singapore Institute of Manufacturing Technology (SIMTech) A*STAR } \\
\IEEEauthorrefmark{3} the Department of Electrical Engineering and Computer
Science, York University\\
\IEEEauthorrefmark{4} Vidyasirimedhi Institute of Science and Technology
\vspace{-5mm}	}

\maketitle\thispagestyle{empty}

\begin{abstract}
Recently, an unmanned aerial vehicle (UAV), as known as drone, has become an alternative means of package delivery. Although the drone delivery scheduling has been studied in recent years, most existing models are formulated as a single objective optimization problem. However, in practice, the drone delivery scheduling has multiple objectives that the shipper has to achieve. Moreover, drone delivery typically faces with unexpected events, e.g., breakdown or unable to takeoff, that can significantly affect the scheduling problem. Therefore, in this paper, we propose a multi-objective and three-stage stochastic optimization model for the drone delivery scheduling, called multi-objective optimization for drone delivery (MODD) system. To handle the the multi-objective optimization in the MODD system, we apply $\varepsilon$-constraint method. The performance evaluation is performed by using a real dataset from Singapore delivery services. 

\end{abstract}
\section{Introduction}

Unmanned aerial vehicles (UAVs), also known as drones, are aerial vehicles that can fly autonomously or be piloted remotely. Thanks to the today's technology, drones are more reliable, efficient, and consume less energy/fuel than before. Business Insider Intelligence has predicted that the sales of drones will reach US\$12 billion by 2021, which is up by a compound annual growth rate (CAGR) of 7.6\%~\cite{ref_bi}. Recently, drones have been used in many industries. Especially, some of major companies have started using drones for delivering parcels for their customers, e.g. Amazon, DHL, Alibaba, and Japan Post. While drones promise to give a cheaper delivery cost, use less manpower, and be more environment-friendly than ground-based vehicles, they have limits on a flying distance and a small carrying capacity. Additionally, a problematic event, e.g. raining and accident, is more likely to occur with adverse effects in drones more than ground-based vehicles. To handle the packages and parcels that cannot be delivered by drones, a shipper may outsource those packages to a carrier. The carrier charges the shipper based on the number of packages, their weights and sizes, etc. The shipper is required to evaluate the delivery plan carefully, as outsourcing the carrier is normally more expensive than using drones. To schedule the package delivery plan, the shipper has multiple objectives to fulfill. On one hand, the shipper wants to obtain a high profit. On the other hand, the shipper is required to provide reliable services and achieve a high satisfaction from customers. 
 
To address the aforementioned challenges, we propose a multi-objective optimization for drone delivery (MODD) system, which aims to help the shipper schedules and plans its package delivery. The objectives are (i) to minimize the total delivery cost, (ii) to minimize the percentage of unsuccessful delivered packages, and (iii) to maximize the reward of on-time delivery. For (ii), the unsuccessful delivery occurs when the drone is unable to take off from the depot or the drone breaks down during the delivery. For (iii), the customers can have one or more specific preferred time slots for delivering and receiving their packages. The time slots are then associated with different rewards to be optimized by the shipper. Moreover, the optimization is formulated as a three-stage stochastic programming to handle the uncertainties of the problematic events, i.e., takeoff condition and breakdown condition. Then, we use $\varepsilon$-constraint method to obtain the exact solutions. Finally, the performance evaluation of the MODD system is presented. The real customer dataset of an industry in Singapore is used in the experiments.

\section{Related Work}

Although there have been a lot of studied on the drone delivery problem~\cite{drone_delivery}~\cite{maggie_tits}~\cite{maggie_vtcfall2018}, they are commonly modeled as a single-objective optimization, despite the fact that industries may consider the drone delivery problem as a multi-objective problem in nature. Multi-objective models for the vehicle routing problem (VRP) has been introduced, and the review of multi-objective VRP can be found in~\cite{multi-obj_vrp}. However, to the best of our knowledge, only the authors in~\cite{multi_d_1},~\cite{multi_d_2}, and~\cite{multi_d_3} considered the multi-objective problem for the drone delivery. The difference of the studies in~\cite{multi_d_1},~\cite{multi_d_2},~\cite{multi_d_3}, and the proposed MODD are summarized in Table 1.
\begin{table*}[t]
\scriptsize
\begin{tabular}{lp{8em}p{8em}p{8em}p{12em}p{25em}}
\hline
\multicolumn{1}{c}{\textbf{}}  & {\textbf{Multi-objective}} & \multicolumn{1}{c}{\textbf{Solver}} & \multicolumn{1}{c}{\textbf{Time involve}} & \multicolumn{1}{c}{\textbf{Stochastic event}}    & \multicolumn{1}{c}{\textbf{Application}}                                                                                                                     \\\hline
{[}6{]}                        & multi-criteria decision making                              & gird based search algorithm (A*)    & min time to destination             & \multicolumn{1}{c}{-}                                                                                               & UAV Parcel delivery.                                                                                                                                          \\ \hline
{[}7{]}                        &$\epsilon$-constraint                                                  & MILP CPLEX                          & time window constraints                   & \multicolumn{1}{c}{-}       & VRP for UAV, but the model is dedicated for VRP with set of Point of Interests (PoIs), i.e., data collection. \\\hline
{[}8{]}                       & multi-objective smart pool search                           & MILP CPLEX                          & min time to destination             & \multicolumn{1}{c}{-}                                                & UAV Parcel delivery. The location and route are considered as x and y coordinates.                                                                            \\\hline
MODD                      & $\epsilon$-constraint                                                  & MILP CPLEX                          & time window constraints                   & take-off and breakdown conditions are considered & UAV Parcel delivery. We consider locations and route similar to traditional VRP, which distance matrix is given.                                                        
\\\hline
\end{tabular}
\caption{The comparison of the related work and our study}

\end{table*}
 The proposed system in~\cite{multi_d_2} addresses the case that drones can visit multiple locations before returning to the depot.  However, in reality, a drone is likely to carry only one package at a time. Therefore, we reformulate the problem for the case that a drone carries one package to serve a customer and returns to the depot before serving the next customer. 
The authors in~\cite{multi_d_3} converted  multi-objectives  into single-objective by adding weight parameters and using the multi-objective smart pool search to adjust the weight parameters. Instead of using the heuristic method as in~\cite{multi_d_3}, we use the $\varepsilon$-constraints to perform the experiments with exact solutions.
Moreover, the drones are less reliable than ground-based vehicles. Therefore, in this paper, we propose a three-objectives and three-stages stochastic optimization to handle random parameters, i.e., takeoff and breakdown events. 
\section{System Model and Assumptions}

In this section, we describe the MODD system, which is formulated as a three-objective optimization. The objectives are (i) to minimize the total delivery cost, (ii) to minimize the percentage of unsuccessful delivered packages, and (iii) to maximize the reward of on-time delivery. The MODD system is formulated as a three-stage stochastic programming. In the system, we consider two types of uncertainty scenarios, i.e., takeoff condition scenario and breakdown condition scenario. We adopt the takeoff and breakdown condition scenarios from~\cite{maggie_tits}. The first-stage and the second-stage are separated by the observation of takeoff condition, and the observation of breakdown condition divides the second-stage and the third-stage from each other. The decisions, which are made in each stage, are as follows: 
\begin{itemize}
\small
\item \textbf{First-stage:} Before the takeoff condition is observed, the drones are reserved, and customers are assigned to either one of the drones or outsourced to a carrier. 
\item \textbf{Second-stage:} The takeoff condition scenario is observed. If the drone can take off, it will deliver the package from the depot to the customer location and will return to the depot. If the drone cannot take off, all packages assigned to the drone are considered to be the unsuccessful deliveries. 

\item \textbf{Third-stage:} After the breakdown condition is observed, if the drone breaks down, the package in the broken drone is regarded as the unsuccessful delivery. Additionally, the packages of the customers that will be served after the breakdown occurs are also regarded as the unsuccessful deliveries. 
\end{itemize} 

The shipper has a set of customers to serve, which is denoted as $\mathcal{C} = \{c_1,c_2,\dots,c_{c'}\}$, where $c'$ represents the total number of customers. We use $i$ and $j$ as indexes of set $\mathcal{C}$. Without loss of generality, each customer has only one package, and the weight of a package of customer $i$ is denoted as $a_i$ kg. The shipper can deliver customers' packages by its drones or outsource the customers' packages to a carrier. Let $\mathcal{D} = \{d_1,d_2,\dots,d_{d'}\}$ denote the set of drones, where $d'$ is the total number of the drones. Each drone has its capacity limit ($g_d$), flying distance limit per trip ($e_d$), flying distance limit per day ($l_d$), start flying time ($\widehat{h}_d$), end flying time ($\bar{h}_d$), and average flying speed ($s_d$). 
If the shipper decides to serve a customer by the drone, the customer also has time preferences that he/she wants the delivery to be done in specific time slots, which can be referred to as a time window. Let $\mathcal{F} = \{f_1,f_2,\dots,f_{f'}\}$ denote the set of time windows, where $f_1$ and $f_2$ represent the most and the second-most preferred time windows, respectively, and $f'$ represents the least preferred time window. Let $\widehat{\mathbb{T}}^{(f)}_i$ and $\bar{\mathbb{T}}^{(f)}_i$ denote the start and the end of the time window that customer $i$ prefers as the $f^{\text{th}}$ order. Let $t_i$ denote the time that a drone needs to spend while serving customer $i$, which can be referred to as serving time.

Moreover, the shipper has a set of depots, i.e., $\mathcal{P} = \{p_1,p_2,\dots,p_{p'}\}$, where $p'$ represents the total number of the depots. Each drone can fly from and return to only one depot. Before delivering, customers' packages can be transferred from an original depot to a new depot, and thus a drone can take the package from the new depot instead of the original depot. Let $o_{i,p}$ be a parameter where $o_{i,p} =1$ when the package of customer $i$ belongs to depot $p$, and $o_{i,p} = 0$ otherwise. We then have the condition $\sum_{p \in \mathcal{P}}o_{i,p} = 1$ for all $i \in \mathcal{C}$. The flying distance from location $u$ to location $v$ is denoted as $k_{u,v}$, where $u$ and $v$ are indexes of a set of locations ($\mathcal{C} \cup \mathcal{P}$).
Let $\Omega = \{\omega_1, \omega_2,\dots,\omega_{\omega'}\}$ be a set of takeoff condition scenarios, where $\omega'$ represents the total number of scenarios in the set. Each $\omega$ is defined as $\omega = \{\mathbb{R}_1,\mathbb{R}_2,\dots,\mathbb{R}_{d'} \}$, where the subscript indicates the drone identification. Again, $d'$ represents the total number of drones. $\mathbb{R}_d = 1$ when drone $d$ cannot take off from the depot, and $\mathbb{R}_d =0$ otherwise. Let $ \Lambda = \{\lambda_1,\lambda_2,\dots,\lambda_{\lambda'}\}$ be a set of breakdown condition scenarios, where $\lambda'$ denotes the total number of the scenarios. Each $\lambda$ is a parameter matrix of $\mathbb{B}_{i,d}$, where $\mathbb{B}_{i,d} = 1$ when drone $d$ breaks down while serving customer $i$, and $\mathbb{B}_{i,d} = 0$ otherwise.

To minimize the total delivery cost, which is one of the objectives of the MODD system, we consider four costs including (i) the initial cost of drones, i.e., $\mathfrak{C}^{(i)}$, (ii) the routing cost, i.e., $\mathfrak{C}^{(r)}$, (iii) the package transfer cost from an original depot to the new depot, i.e., $\mathfrak{C}^{(t)}$, and (iv) the outsourcing cost, i.e., $\mathfrak{C}^{(c)}$. 


\section{Multi-Objective optimization}

In this section, we present the optimization problem formulations of the MODD system. The detail of the three objective functions and method to solve the multi-objective optimization problem are presented in Section~\ref{sec_obj}. We define the decision variables and the constraints of the MODD system in Section~\ref{sec_decision} and~\ref{sec_constraints}, respectively.

\subsection{Multiple objective functions}
\label{sec_obj}

There are three objective functions in the MODD system, i.e., (i) to minimize the total delivery cost, (ii) to minimize the percentage of unsuccessful delivered packages, and (iii) to maximize the reward of on-time delivery. The formulation of these objectives are presented in~(\ref{eq_obj_cost}),~(\ref{eq_obj_break}), and~(\ref{eq_obj_schedule}), respectively. $\mathbb{O}^{\mathrm{C}}$ represents the total delivery cost, $\mathbb{O}^{\mathrm{U}}$ represents the percentage of unsuccessful delivered packages, and $\mathbb{O}^{\mathrm{R}}$ represents the reward of on-time delivery. 


\begin{myalign}
&\text{Minimize: }\mathbb{O}^{\mathrm{C}}=&\sum_{d\in \mathcal{D}} \mathfrak{C}^{(i)}_dW_{d} 
+ \hspace{-2em}\sum_{i \in \mathcal{C}, d\in \mathcal{D}, p \in \mathcal{S}}\hspace{-2em}\left( \mathfrak{C}^{(r)}_{p,i} + \mathfrak{C}^{(r)}_{i,p} \right) Y_{i,d,p} \nonumber \\ 
&&+ \sum_{p \in \mathcal{S}}\mathfrak{C}^{(t)}_{p}T_p + \sum_{i \in \mathcal{C}}\mathfrak{C}^{(c)}_iZ_i 
\label{eq_obj_cost}
\end{myalign}

\begin{myalign}
&\text{Minimize: }\mathbb{O}^{\mathrm{U}}= 
\frac{100}{c'}\sum_{\substack{\omega \in \Omega\\ i \in \mathcal{C}\\ d \in \mathcal{D}}} \left( \mathbb{P}(\omega)X^{(b)}_{i,d}(\omega) + \sum_{\lambda \in \Lambda} \mathbb{P}(\omega, \lambda)X^{(a)}_{i,d}(\omega,\lambda) \right)
\label{eq_obj_break}
\end{myalign}
\vspace{-1.5em}

\begin{myalign}
\text{Maximize: }\mathbb{O}^{\mathrm{R}} = 
\sum_{i \in \mathcal{C}, f \in \mathcal{F}} C^{(f)}F^{(f)}_i 
\label{eq_obj_schedule}
\end{myalign}
Note that $C^{(f)}$ represents the constant parameter, where $C^{(f_1)} > C^{(f_2)} > \dots > C^{(f_{f'})}$. Again, $f_1$ is the most preferred time window and $f_{f'}$ is the least preferred time window.

To solve the problem, we convert the proposed multi-objective optimization to a linear programming by the $\varepsilon$-constraint method~\cite{multi_d_2}. We can use one of the three objectives in (\ref{eq_obj_cost}) to (\ref{eq_obj_schedule}) as the objective function and use the others as constraints. An example of the conversion are as follows:

\noindent Minimize: $\mathbb{O}^{\textrm{C}}$\\
 subject to (\ref{eq_multi_1}), (\ref{eq_multi_2}), and (\ref{eq_con_takeoff}) to (\ref{eq_con_schedule}).

\begin{myalign}
\mathbb{O}^{\textrm{U}} \leq \epsilon_1 \label{eq_multi_1}\\
\mathbb{O}^{\textrm{R}} \geq \epsilon_2 \label{eq_multi_2}	.
\end{myalign}

The effective solutions of the MODD system can be achieved by vary the parameters $\epsilon_1$ and $\epsilon_2$. Again, the objective function can be to minimize $\mathbb{O}^{\textrm{C}}$, to minimize $\mathbb{O}^{\textrm{U}}$, or to maximize $\mathbb{O}^{\textrm{R}}$. Once the objective is selected, the other objectives are taken as the constraints with parameters $\epsilon_k$. The steps of solving the multi-objective optimization by $\varepsilon$-constraints are listed below.
\begin{description}\small
\item[Step 1] Initialize the range of $\epsilon_k$ parameters~\cite{ref_eps}.
\item[Step 2] Solve the linear programming problem by using parameter $\epsilon$ with the smallest value.
\item[Step 3] Increase the value of one of the parameters $\epsilon_k$ and repeat Step 2. Once the value of parameter $\epsilon_k$ is not in the range, the algorithm is terminated. 
\item[Step 4] Obtain the set of solutions. 
\end{description}

Note that we do not select the best solution for the shipper because the different shipper may have different criterion for the multi-objective optimization. In this paper, we present the feasible solutions for the shipper to select. 
\vspace{-1em}

\subsection{Decision Variables}
\label{sec_decision}

There are thirteen decision variables in the MODD system. All the decision variables are binary, except $U_{i,d}$ which is an integer, and $Q_i$ which is a positive variable. The definitions of the decision variables are as follows.

\begin{itemize}
\small
\item $W_{d}$ is the indicator for determining whether drone $d$ is used or not, i.e., if $W_{d} =1$, drone $d$ will be used in the delivery, and $W_{d}=0$ otherwise. 
\item $Y_{i,d,p}$ is the allocation variable. If $Y_{i,d,p} =1$, customer $i$ will be served by drone~$d$, and the drone will depart from depot~$p$, and $Y_{i,d,p} =0$ otherwise.
\item $Z_i$ is the indicator for determining whether  the package of customer~$i$ will be delivered by a carrier or not. If $Z_i =1$, customer $i$ will be served by the outsourcing carrier, and $Z_i=0$ otherwise. 
\item $T_p$ is the indicator for determining whether whether the shipper has to transfer packages from/to depot $p$ or not. $T_p =1$ means that at least one package is transferred from/to depot $p$, and $T_p =0$ otherwise. 
\item $M_{i,p,q}$ is the indicator for determining  whether the package of customer~$i$ is transferred from depot $p$ to depot~$q$ or not. If $M_{i,p,q}=1$, the package of customer~$i$ is transferred from depot~$p$ to depot~$q$, and $M_{i,p,q}=0$ otherwise. 
\item $B_{d,p}$ is an auxiliary variable for imposing the drone to have only one departing and returning depot. 
\item $U_{i,d}$ is a serving order of drone $d$, where $U_{i,d} < U_{j,d}$ means that drone $d$ will serve customer $i$ before customer $j$.
\item $A_{i,j,d}$ is an auxiliary variable for ensuring that two customers cannot have the same serving order when they are served by the same drone. 
\item $X_{i,d}^{(b)}(\omega)$ is the takeoff condition variable in which $X_{i,d}^{(b)}(\omega)= 1$ means that drone $d$ cannot takeoff to serve customer $i$ under scenario $\omega$, and $X_{i,d}^{(b)}(\omega)= 0$ otherwise. 
\item $X_{i,d}^{(a)}(\omega,\lambda)$ is the breakdown condition variable in which $X_{i,d}^{(a)}(\omega,\lambda) = 1$ means that drone $d$ breaks down during serving customer $i$ under takeoff scenario $\omega$ and breakdown scenario $\lambda$. Otherwise, $X_{i,d}^{(a)}(\omega,\lambda) = 0$.
\item $F_i^{(f)}$ is the indicator for determining whether the $f^{\text{th}}$ preferred time window is selected or not. $F_i^{(f)} =1$ means that customer $i$ will be served during the time window that the customer prefers as the $f^{\text{th}}$ order, and $F_i^{(f)} =0$ otherwise.
\item $V_{i,j,d}$ is the indicator for determining whether drone $d$ serves customer $i$, and then it serves customer $j$ as the next customer or not. If $V_{i,j,d}=1$, drone $d$ will serve customer $i$ before customer $j$, and $V_{i,j,d}=0$ otherwise.
\item $Q_i$ is a serving time variable. If $Q_i \ge Q_j$, customer $i$ will be served before customer $j$. 
\end{itemize}

\begin{figure*}
\vspace{-1em}
\begin{minipage}{0.65\textwidth}
\centering
\includegraphics[trim={0 0 0 1cm},width=0.9\textwidth, height =11em]{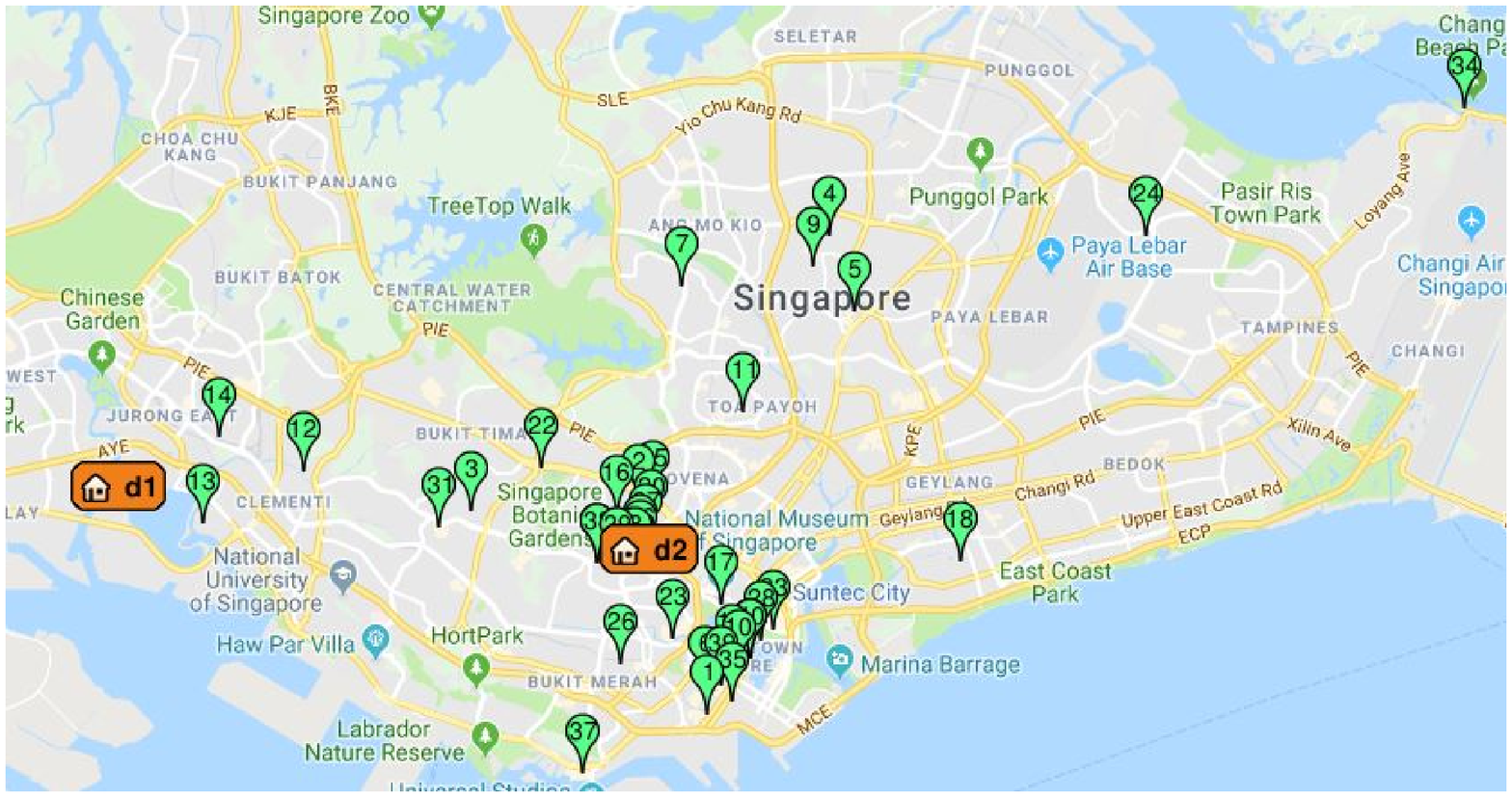}
\caption{Map}
\end{minipage}\begin{minipage}{0.3\textwidth}
\includegraphics[trim={0.5cm 0 0.5cm 1cm},width=\textwidth]{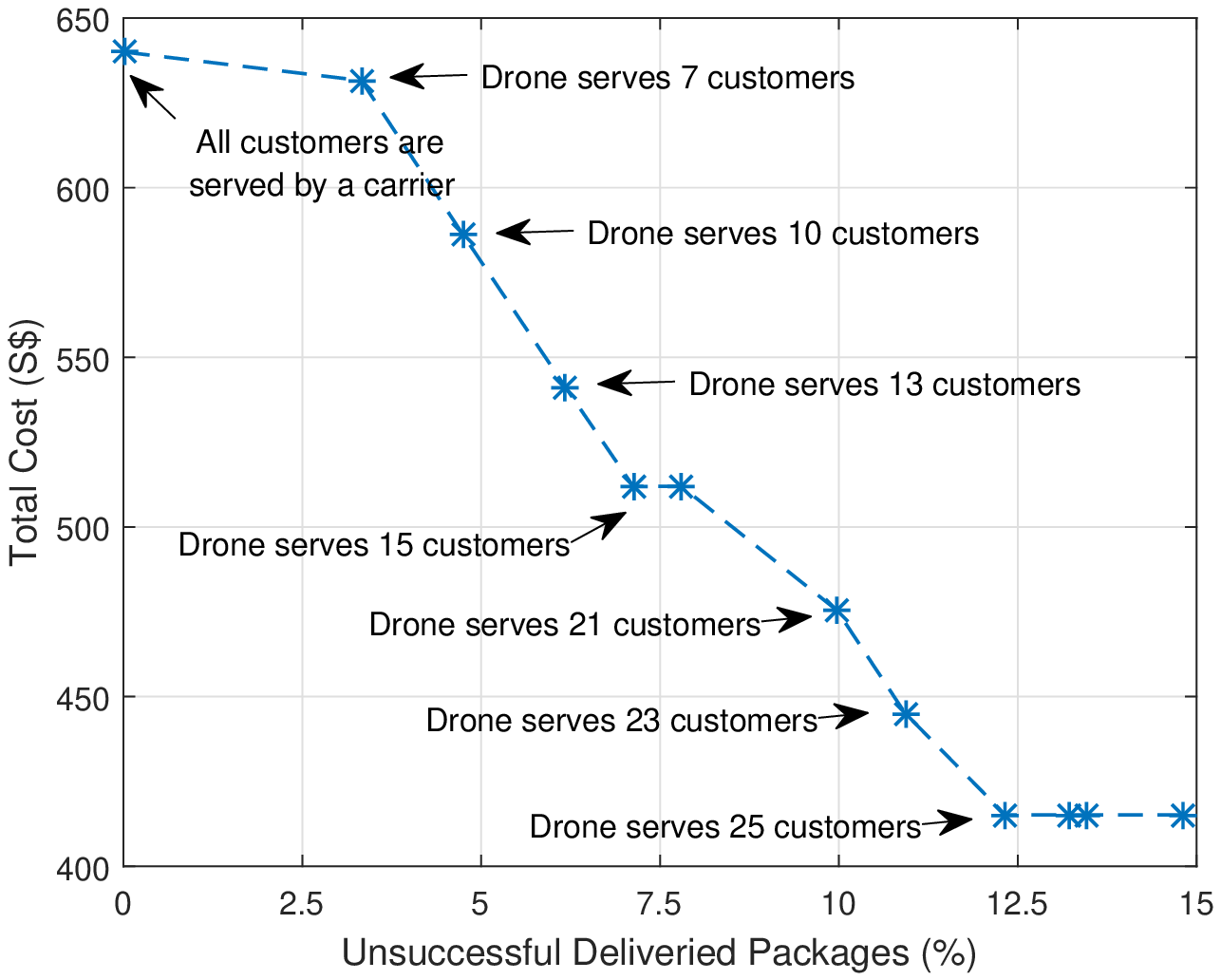}
\caption{The Pareto frontier}
\label{fig_cost_ave}
\end{minipage}
\vspace{-1em}
\end{figure*}

\subsection{Constraints}
\label{sec_constraints}

There are four groups of the constraints in the MODD system including (i) general constraints for the drone delivery with package transfer, (ii) breakdown constraints for the drone delivery, (iii) the serving order constraints for the drone delivery, and (iv) time window constraints with the reward counter. The general constraints for the drone delivery that we use in this paper are similar to the constraints in (2), (3), (5), and (7) to (16) of~\cite{maggie_vtcfall2018}. The constraints include the initial cost constraint, package allocation constraint, package transfer constraints, traveling time limit constraints, traveling distance limit constraints, and capacity constraint. The formulations and the explanations of (ii), (iii), and (iv) are presented as follows:

\begin{myalign}
&\sum_{p \in \mathcal{S}} Y_{i,d,p} \mathbb{R}_d(\omega) - Z_{i} = X^{(b)}_{i,d}(\omega) ,\hspace{1em}\forall i \in \mathcal{C}, d \in \mathcal{D}, \omega \in \Omega 
\label{eq_con_takeoff}\\
 &\sum_{p \in \mathcal{S}} Y_{i,d,p} \left(1 -\mathbb{R}_d(\omega)\right) \mathbb{B}_{i,d}(\lambda) \leq X^{(a)}_{i,d}(\omega,\lambda), \nonumber \\
 & \hspace{11em}\forall i \in \mathcal{C}, d \in \mathcal{D}, \omega \in \Omega, \lambda \in \Lambda \label{eq_con_break1}\\
& U_{i,d} - U_{j,d} \leq \Delta\left(1-X^{(a)}_{j,d}(\omega,\lambda) + X^{(a)}_{i,d}(\omega,\lambda) \right),
\nonumber \\ 
& \hspace{8em}i \neq j, \forall i,j \in \mathcal{C}, d \in \mathcal{D}, \omega \in \Omega, \lambda \in \Lambda \label{eq_con_break2}
\end{myalign}

\begin{myalign}
&\sum_{p \in \mathcal{S}}Y_{i,d,p} \leq U_{i,d}, 
\hspace{10em} \forall i \in \mathcal{C}, d \in \mathcal{D} \label{eq_con_order1}\\
&0 \leq U_{j,d} \leq \sum_{i \in \mathcal{C}, p \in \mathcal{S}}Y_{i,d,p}, \hspace{6em}\forall j \in \mathcal{C}, d \in \mathcal{D} \label{eq_con_order2}\\
& U_{i,d} - U_{j,d} \leq \Delta A_{i,j,d} - \sum_{p \in \mathcal{S}}Y_{i,d,p}, \nonumber \\
& \hspace{14em} i \neq j, \forall i,j \in \mathcal{C}, d \in \mathcal{D} \label{eq_con_order3}\\
& U_{i,d} - U_{j,d} \geq \sum_{p \in \mathcal{S}}Y_{i,d,p} - \Delta (1-A_{i,j,d}) , \nonumber \\
&\hspace{14em} i \neq j, \forall i,j \in \mathcal{C}, d \in \mathcal{D} \label{eq_con_order4}\\
& U_{j,d} - U_{i,d} \leq \Delta V_{i,j,d}, \hspace{4em}i\neq j,\forall i,j \in \mathcal{C}, d \in \mathcal{D} \label{eq_con_order5} 
\end{myalign}

The constraints in (\ref{eq_con_takeoff}) and (\ref{eq_con_break1}) ensure that the percentage of unsuccessful delivered packages is calculated from the number of packages that cannot be delivered when the drone cannot take off and when the drone breaks down during serving, respectively. The constraint in (\ref{eq_con_break2}) ensures that the rest of the packages are taken into account in the number of unsuccessful delivered packages after the drone breaks down. For example, $U_{j,d} > U_{i,d}$ and drone $d$ breaks down while serving customer $i$, drone $d$ will not be able to deliver the package of customer~$j$.

The constraints in (\ref{eq_con_order1}) to (\ref{eq_con_order4}) are the serving order constraints of drones. The constraints in (\ref{eq_con_order1}) and (\ref{eq_con_order2}) are the boundary constraints of serving order variables. The constraints in (\ref{eq_con_order3}) and (\ref{eq_con_order4}) ensure that the customers have the different serving order if they are served by the same drone. These constraints are similar to those in~\cite{maggie_tits}.

\begin{myalign}
&Q_i + \sum_{p \in \mathcal{S}} \left( \dfrac{k_{i,p}+k_{p,i}}{s_d} + t_i \right)Y_{i,d,p} - Q_j \leq \Delta(1-V_{i,j,d}),
\nonumber \\ & \hspace{18em} i\neq j, \forall i, j \in \mathcal{C} \label{eq_con_q1}\\
& \widehat{h}_d \leq Q_i +Z_i 
\hspace{4.5em}\forall i \in \mathcal{C}, d \in \mathcal{D} \label{eq_con_q1.1}\\
& Q_i \leq \bar{h}_d - \frac{k_{i,p}+k_{p,i}}{s_d} - t_i, 
\hspace{4.5em}\forall i \in \mathcal{C}, d \in \mathcal{D} \label{eq_con_q2}\\
&\widehat{\mathbb{T}}^{(f)}_i - Q_i + \sum_{p,d}\dfrac{k_{i,p}+k_{p,i}}{s_d}Y_{i,d,p} \leq \Delta(1-F_i^{(f)}),\nonumber \\
 &\hspace{18.5em} \forall i \in \mathcal{C}, f \in \mathcal{F} \label{eq_con_q3}
\\
&Q_i + \sum_{p,d}\left(\dfrac{k_{i,p}+k_{p,i}}{s_d}+t_i \right) Y_{i,d,p} - \bar{\mathbb{T}}^{(f)}_i \leq \Delta(1-F_i^{(f)}), 
\nonumber \\& \hspace{18.5em}\forall i \in \mathcal{C}, f \in \mathcal{F} \label{eq_con_q4}\\
&\sum_{f \in \mathcal{F}}F_i^{(f)} \leq 1, \hspace{5em} \forall i \in \mathcal{C} \label{eq_con_schedule}
\end{myalign}

The constraints in (\ref{eq_con_order5}) to (\ref{eq_con_q2}) ensure that (i) serving time $Q_i < Q_j$ with the serving order $U_i < U_j$, and (ii) the time between $Q_i$ and $Q_j$ must be longer than the time of flying from customer $i$ to the depot plus flying from the depot to customer $j$. The constraints in (\ref{eq_con_q3}) and (\ref{eq_con_q4}) ensure that serving time $Q_i$ is between the period of the selecting time window. For example, if the most preferred time window ($f=1$) of customer $i$ is selected, i.e., $F^{(1)}_i =1$, then $ \widehat{\mathbb{T}}_i^{(1)}\leq Q_1\leq \bar{\mathbb{T}}_i^{(1)}$. Again, $ \widehat{\mathbb{T}}_i^{(1)}$ and $\bar{\mathbb{T}}_i^{(1)}$ are the start and the end of the time window. The constraint in (\ref{eq_con_schedule}) ensures that only one time window can be selected.

Next, we evaluate the MODD system and present the experimental results. 


\section{Performance Evaluation}

In this section, we evaluate the MODD system with the real customer data from a Singapore logistic company. Forty customers are considered in all the experiments. Each customer has a location, the start of time window, and the end of time window. We assume that all the packages of customers are $a_i \leq 5$ kg. Therefore, the cost of outsourcing package to a carrier is set as $\mathfrak{C}^{(c)} = S\$16$ based on the Speedpost service of SingPost company. 
The shipper has two depots, i.e., $\mathcal{P}=\{p_1,p_2\}$, and two drones, i.e., $\mathcal{D}=\{d_1,d_2\}$. The drones are of the same type, where $\mathfrak{C}^{(i)}_d = S\$100$, $l_d =150$ km, $e_d=10$ km, $g_d= 5$ kg, $h_d= 8$ hrs, and $s_d = 30$ km/hr. The package transferring cost of both depots is set as $\mathfrak{C}^{(t)}= S\$30$. We assume that routing of drones is similar to ground-based vehicle because Singapore has many high buildings, e.g., residential housing. The routing cost can be calculated by $\mathfrak{C}^{(r)} = distance \times 1.05 \times 0.1 $, where the distance in kilometers between one location and the other location is extracted from Google Map. Two takeoff and breakdown scenarios are considered in all the experiments, i.e., $\Lambda = \{\lambda_1,\lambda_2\}$ and $\Omega = \{\omega_1,\omega_2\}$. Let $\lambda_1$ be the scenario that all the drones can take off and $\lambda_2$ be the scenario that all the drones cannot take off, e.g. due to raining. Let $\omega_1$ be the scenario that breakdown does not occur and $\omega_2$ be the scenario that breakdown occurs every time when the drones serve customers. The probabilities of the scenarios are set as $P(\lambda_1)=P(\omega_1)=0.9$ and $P(\lambda_2)=P(\omega_2)=0.1$. Note that the probabilities can be calculated based on the history record. 
We implement the GAMS script for the optimization problem and solve it by the solver CPLEX~\cite{ref_gams}.

 \begin{figure*}
 \vspace{-1.5em}
\begin{minipage}{0.63\textwidth}
\centering\scriptsize
$\begin{array}{cc}
\includegraphics[trim={0.5cm 0 0.5cm 0.7cm},clip,width=0.5\textwidth]{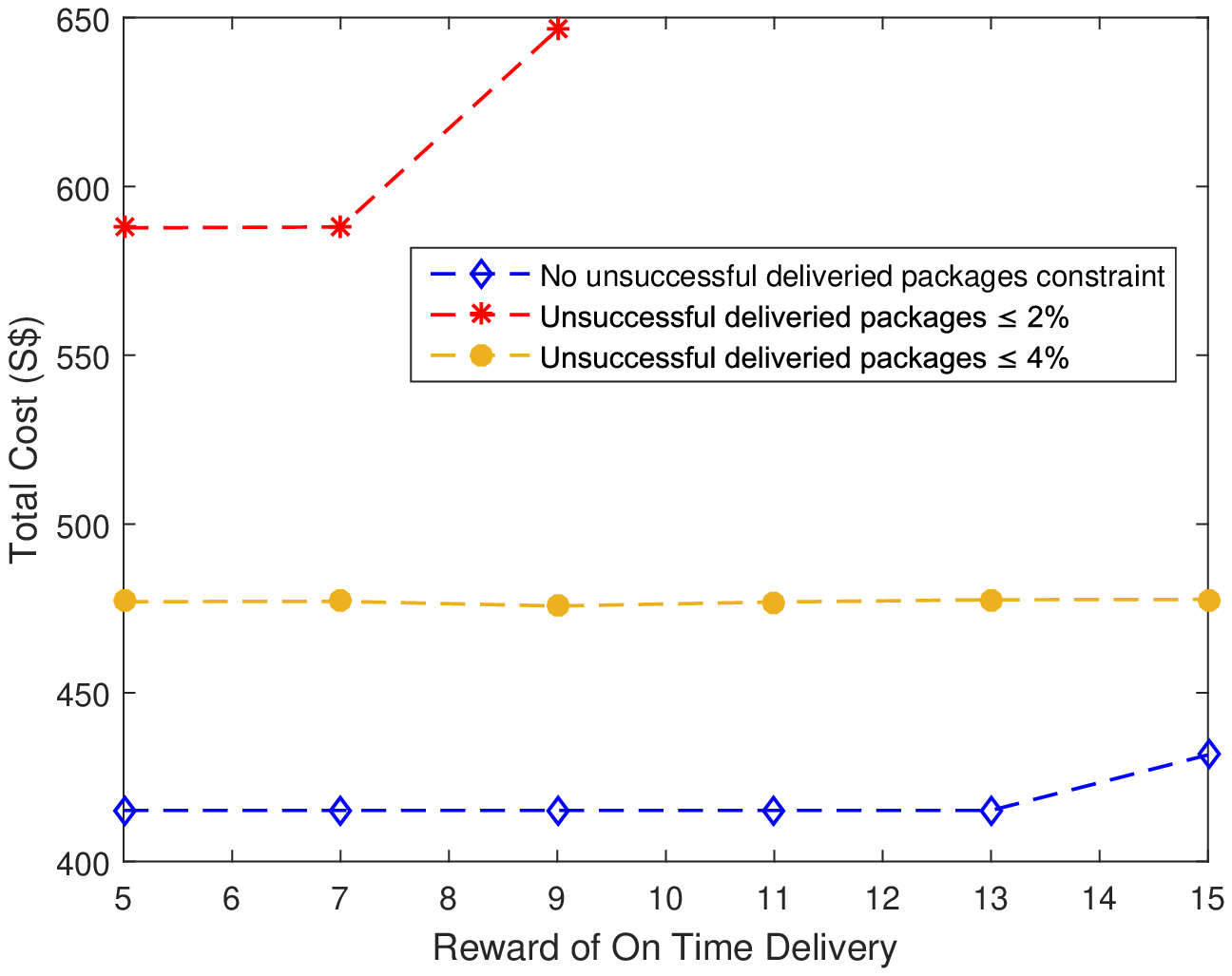} & 
\includegraphics[trim={0.5cm 0 0.3cm 1cm},width=0.5\textwidth]{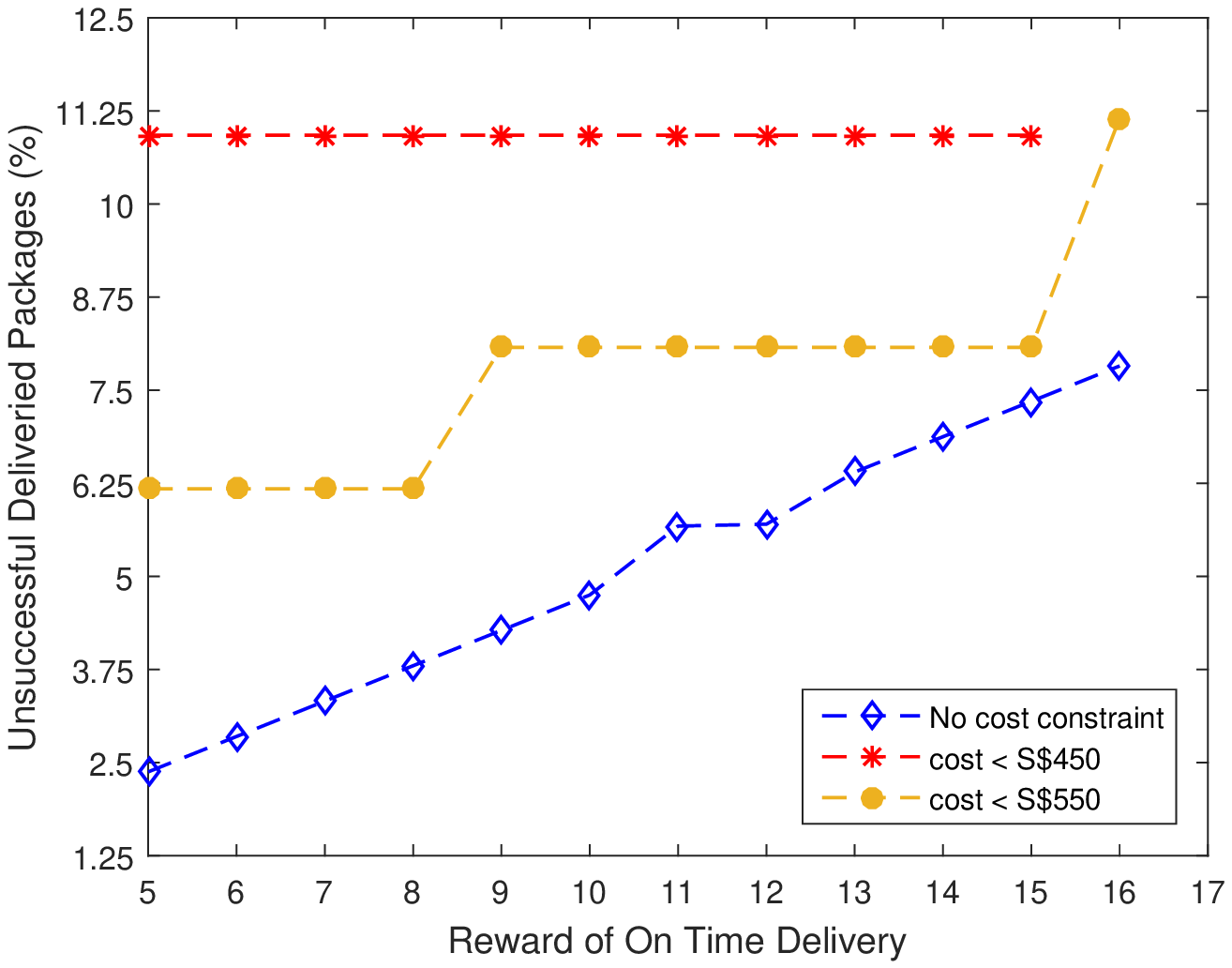}\\
 (a) & (b)\end{array}$
\caption{Impact of number of matching required time windows while minimizing $(a)$ the total cost and $(b)$ the percentage of unsuccessful delivered packages.}
\label{fig_varyRequired}
\end{minipage}\hspace{3em}\begin{minipage}{0.3\textwidth}
\centering
\includegraphics[trim={0.5cm 0 0.5cm 1cm},width=\textwidth]{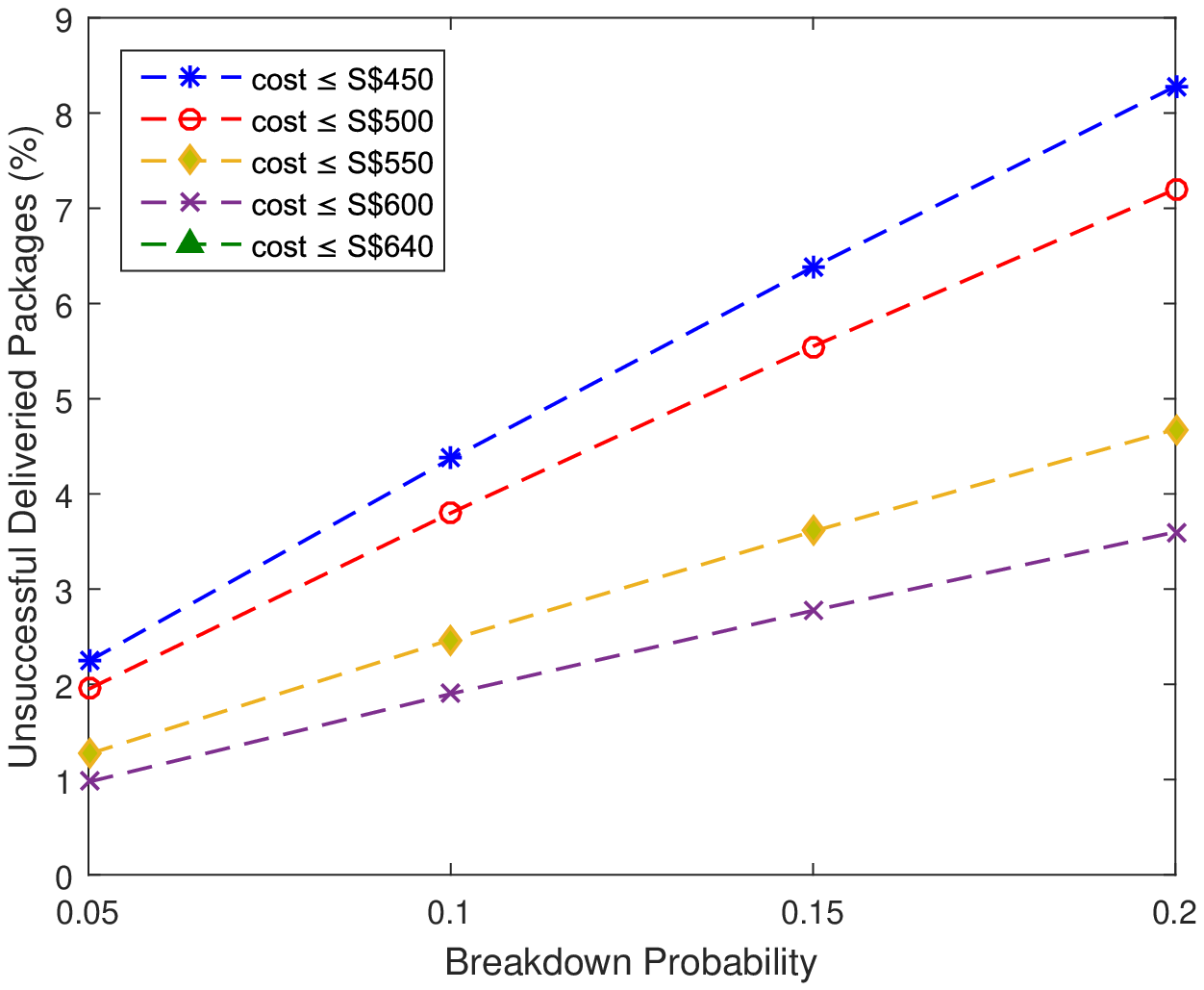}
\caption{The impact of breakdown probability on the percentage of unsuccessful delivered packages.}
\label{fig_varyProb}
\end{minipage}
\vspace{-1em}
\end{figure*}

\vspace{-1em}

\subsection{Pareto frontier}

We experiment the system with two objectives, which are to minimize the total cost, i.e., $\mathbb{O}^{\mathrm{C}}$, and to minimize the percentage of unsuccessful delivered packages, i.e., $\mathbb{O}^{\mathrm{U}}$. The Pareto frontier of these two objectives is presented in Figure~\ref{fig_cost_ave}. To guarantee that the shipper will not fail to deliver any packages, the shipper can outsource all packages to the carrier, which is more expensive than delivering by using the shipper's own drones. The total cost is stable when the percentage of unsuccessful delivered packages is more than $13.325\%$. For this experiment setting, the shipper needs to pay at least $S\$415.16$ to deliver all packages.

\vspace{-1em}

\subsection{Reward of on-time delivery}

To demonstrate the impact of the reward of on-time delivery, we present two test cases, i.e., (i) minimizing the total delivery cost in Figure~\ref{fig_varyRequired}(a) and (ii) minimizing the percentage of unsuccessful delivered packages in Figure~\ref{fig_varyRequired}(b). Note that if we use (i) or (ii) as an objective function, then (ii) and (i) will become a constraint, respectively.

From Figure~\ref{fig_varyRequired}(a), we reach fifteen as the highest number of matching time windows by using two drones to serve all the customers in the coverage area of the drones. When the reward of on-time delivery is larger than or equal to 17, the shipper has not enough drones to satisfy the time window requirement of all the customers. The total cost is higher when we force the percentage of unsuccessful delivered packages to be lower than a certain value as the shipper needs to outsource some packages to reduce the breakdown event. 

From Figure~\ref{fig_varyRequired}(b), to minimize the percentage of unsuccessful delivered packages, the shipper can outsource as many packages as possible to the carrier. Consequently, these packages will be delivered without experiencing the drone breakdown. The shipper can use one drone to serve the customers when the total cost needs to be lower than S\$450. When the total cost constraint is less than or equal to $S\$450$, the percentage of unsuccessful delivered packages is higher than that when the total cost constraint is less than or equal to $S\$550$. The reason is that the number of outsourced packages is fewer. If the shipper does not consider the total cost constraint, the percentage of unsuccessful delivered packages varies linearly with respect to the reward of on-time delivery.

\vspace{-1em}

\subsection{Impact of probabilities ($\lambda$ and $\omega$)}

We set the takeoff conditional probability equal to the breakdown condition probability, i.e., $P(\lambda_1)=P(\omega_1)$ and vary them. For ease of the presentation, we do not consider the reward of on-time delivery in this experiment. 


When the total cost is less than or equal to $S\$450$, 23 customers are served by a drone, and the rest are served by the carrier. Similarly, 20 customers, 27 customers, and 30 customers are served by the drone when the total cost is less than or equal to $S\$500$, $S\$550$, and $S\$600$, respectively. From Figure~\ref{fig_varyProb}, we can conclude that the percentage of unsuccessful delivered packages varies proportionally to the breakdown probability, except when the total cost is larger than or equal to $S\$640$. When the total cost is larger than or equal to $S\$640$, the shipper will not experience the drone breakdown because the shipper can outsource all packages to the carrier, i.e., $(40 \times S\$16)$, which will not incur the percentage of unsuccessful delivered packages. 

\section{Conclusion}

We have proposed the multi-objective optimization for drone delivery (MODD) system to help the shipper schedule and plan its delivery by providing a set of potential solutions. We have formulated three different objectives in the system, i.e., to minimize the total delivery cost, to minimize the percentage of unsuccessful delivered packages, and to maximize the reward of on-time delivery. The trade-off between using drones and outsourcing packages to a carrier has been considered in the system as drones may not be able to reach some customers, e.g., due to flying distance limit. Furthermore, we have formulated the MODD system as a three-objective and three-stage stochastic programming. The takeoff and breakdown conditions are taken into account during the second and the third stages of the optimization, respectively. We have evaluated the MODD system with the real data from a Singapore company, and we presented the analysis of the Pareto frontiers of the system. 

\vspace{-0.7em}
\section{Acknowledgment}
\scriptsize
This work was partially supported by Singapore Institute of
Manufacturing Technology-Nanyang Technological University
(SIMTech-NTU) Joint Laboratory and Collaborative research
Programme on Complex Systems.


\vspace{-1em}

\begin{thebibliography}{1}

\bibitem{ref_bi} Business Insider. [Online]. Available:https://www.businessinsider.com/com mercial-uav-market-analysis-2017-8/?IR=T

\bibitem{drone_delivery} K. Dorling, J. Heinrichs, G. G. Messier and S. Magierowski, ``Vehicle Routing Problems for Drone Delivery," in {\em IEEE Transactions on Systems, Man, and Cybernetics}, vol. 47, no. 1, pp. 70-85, Jan 2017.

\bibitem{maggie_tits} S. Sawadsitang, D. Niyato, P. Tan and P. Wang, ``Joint Ground and Aerial Package Delivery Services: A Stochastic Optimization Approach," in {\em IEEE Transactions on Intelligent Transportation Systems.}

\bibitem{maggie_vtcfall2018} S. Sawadsitang, D. Niyato, P.S. Tan,P. Wang, ``Supplier Cooperation in Drone Delivery", {\em 2018 IEEE 88th Vehicular Technology Conference (VTC fall)}, Chicago, USA, Aug 2018.



\bibitem{multi-obj_vrp}N. Jozefowiez, F. Semet, and E.-G. Talbi, ``Multi-objective vehicle routing problems," {\em European Journal of Operational Research}, Vol. 189, Issue 2, 2008, PP. 293-309.

\bibitem{multi_d_1} P. Wu and D. A. Campbell, and T. Merz, ``On-board multi-objective mission planning for unmanned aerial vehicles." {\em IEEE Aerospace Conference}, 7-14 March 2009, Big Sky, Montana.

\bibitem{multi_d_2} F. Guerriero, R. Surace, V. Loscrí, E. Natalizio, ``A multi-objective approach for unmanned aerial vehicle routing problem with soft time windows constraints," {\em Applied Mathematical Modelling}, Vol. 38, Issue 3, 2014, PP. 839-852.

\bibitem{multi_d_3} B. N. Coelho, V. N. Coelho, I. M. Coelho, L. S. Ochi, R. Haghnazar K., D. Zuidema, M. S.F. Lima, A. R. Costa, ``A multi-objective green UAV routing problem," {\em Computers \& Operations Research}, Vol. 88, 2017, PP. 306-315.


\bibitem{ref_eps} G. Mavrotas,
``Effective implementation of the ε-constraint method in Multi-Objective Mathematical Programming problems," {\em Applied Mathematics and Computation}, Vol. 213, Issue 2, 2009, PP. 455-465.

\bibitem{ref_singpost}Singpost. [Online]. Available: \url{http://www.singpost.com/}

\bibitem{ref_gams}D. Chattopadhyay, ``Application of General Algebraic Modeling System to Power System Optimization," {\em IEEE Transactions on Power Systems}, vol. 14, no. 1, pp. 15-22, Feb 1999.

\end{thebibliography}
\end{document}